\documentstyle[preprint,aps]{revtex}

\begin{document}
\title{Quantum Entanglement of Excitons in Coupled Quantum Dots}
\author{Ping Zhang$^1$, C. K. Chan$^2$, Qi-Kun Xue$^1$, Xian-Geng Zhao$^3$}
\address{$^1$International Center for Quantum Structure and State Key laboratory for\\
Surface Physics, Institute of Physics, The Chinese Academy of Sciences,\\
Beijing 100080, China\\
$^2$Department of Applied Mathematics, The Hong Kong Polytechnic University,%
\\
Hong Kong \\
$^3$Institute of Applied Physics and Computational Mathematics, P.O.Box\\
8009, Beijing 100088, China}
\date{\today }
\maketitle

\begin{abstract}
Optically-controlled exciton dynamics in coupled quantum dots is studied. We
show that the maximally entangled Bell states and
Greenberger-Horne-Zeilinger (GHZ) states can be robustly generated by
manipulating the system parameters to be at the avoided crossings in the
eigenenergy spectrum. The analysis of population transfer is systematically
carried out using a dressed-state picture. In addition to the quantum dot
configuration that have been discussed by Quiroga and Johnson [Phys. Rev.
Lett. {\bf 83}, 2270 (1999)], we show that the GHZ states also may be
produced in a ray of three quantum dots with a shorter generation time.%
\newline %
\newline %
\end{abstract}

\section{Introduction}

Entanglement is of great interest in many areas of active research in
contemporary quantum physics, such as quantum computation\cite{Ekert},
quantum teleportation\cite{Ben}, and fundamental tests of quantum mechanics 
\cite{Bell,Gree}. How to design and realize quantum entanglement is
extremely challenging due to the intrinsic decoherence, which is caused by
the uncontrollable coupling with environmental degrees of freedom. A variety
of physical systems have be chosen to investigate the controlled, entangled
states. Among these are trapped ions\cite{Cirac}, spins in nuclear magnetic
resonance (NMR)\cite{Ger}, cavity-quantum-electrodynamics systems\cite{Dom},
Josephson junctions\cite{Mak}, and quantum dots\cite{Loss}.

Recently, the combination of progresses in ultrafast optoelectronics\cite
{Shah} and in nanostructure fabrication\cite{Jac} brings out dense study of
the coherent-carrier control in semiconductor quantum dots (QDs). Present
ultrafast laser technology allows the coherent manipulation of carrier
(electron and/or hole) wave functions on a time scale shorter than typical
dephasing times\cite{Bon}. It has been envisioned that optical excitations
in QDs could be successfully exploited for quantum information processing:
Quiroga and Johson\cite{Qui}, and Reina {\it et al}.\cite{Rei1,Rei2}
suggested that the resonant transfer interaction between spatially separated
excitons in quantum dots can be exploited to produce many-particle
entanglement. Based on numerical analysis of realistic double QDs, Biolatti 
{\it et al}.\cite{Bio1}, and Troiani {\it et al.}\cite{Tro} proposed an {\it %
all optical} implementation of quantum information processing. Chen {\it et
al.}\cite{Chen1}, and Piermarocchi {\it et al}.\cite{Pie} suggested the
controlling of spin dynamics of two interacting excitons with pulses of
spin-polarized optical excitations. Stievater {\it et al}.\cite{Sti}
successfully observed the single qubit rotation of excitonic Rabi
oscillation and in a QD. Chen {\it et al}.\cite{Chen2} measured the quantum
entanglement between a pair of electron and hole. Furthermore, Bayand {\it %
et al}.\cite{Bay} demonstrated the entanglement of electron-hole pairs. Up
to now, the basic ingredient double-qubit operation, i.e., the
controlled-not (CNOT) operation has not been experimentally demonstrated.

In this paper we study the optical control of the exciton dynamics in
multiple QDs. Following Ref. [13], we assume that the excitonic occupation
operator $\widehat{n}_l$ for the $l$-th QD has only two eigenvalues $n_l=0$
and $n_l=1$, corresponding to the absence and the presence of a ground-state
exciton. Thus the single-qubit basis consist of $|0\rangle _l$ and $%
|1\rangle _l$. The whole computational state space is spanned by the basis $|%
{\bf n\rangle =\otimes }_l|n_l\rangle $ ($n_l=0,1$). We show that the
avoided crossing in eigenenergy spectrum enables the robust generation of
maximally entangled Bell state of two qubits, and GHZ states of three
qubits. The entangled state generation time is analytically obtained by
adiabatically eliminating the dark multi-exciton states.

This paper is organized as follows. Section II contains the theoretical
model: in Sec. II A we present the Hamiltonian of the multiple QDs with
equidistance from each other, whereas the Hamiltonian of the QDs with a
linear arrangement is presented in Sec. II B. The maximally entangled Bell
state generation is showed in Sec. III. In Sec. IV the maximally entangled
GHZ state generation is shown for the three QDs with equaidistance. The GHZ
state generation process for the QDs with a linear configuration is analyzed
in Sec. V. A summary is given in Sec. VI.

\section{Theoretical model}

We consider a system of $N$ identical QDs radiated by classical optical
field. Ignoring any constant energy terms, the Hamiltonian describing the
formation of single excitons within the individual QDs and their interdot
hopping is given by 
\begin{eqnarray}
H(t) &=&\frac \varepsilon 2\sum_{n=1}^N(e_n^{+}e_n-h_n^{+}h_n)-\frac 12
\sum_{n,n^{\prime }=1}^NV_{nn^{\prime }}(e_n^{+}h_{n^{\prime }}e_{n^{\prime
}}h_n^{+}+h_ne_{n^{\prime }}^{+}h_{n^{\prime }}^{+}e_n)  \eqnum{1} \\
&&+\frac{\Omega (t)}2e^{-i\omega t}\sum_{n=1}^Ne_n^{+}h_n^{+}+\frac{\Omega
^{*}(t)}2e^{i\omega t}\sum_{n=1}^Nh_ne_n  \nonumber
\end{eqnarray}
in the rotating wave approximation. Here $e_n^{+}$ ($h_n^{+}$) is the
electron (hole) creation operator in the $n$-th QD, $\varepsilon $ is the QD
band gap while $V_{nn^{\prime }}$ represents the interdot Coulomb
interaction between the $n$-th and $n^{\prime }$-th QDs, the time dependence
of $\Omega (t)$ describes the laser pulse shape while $\omega $ is the
optical frequency. As in the atomic case, the condition $\omega \gg |\Omega
(t)|$ enables the rotating wave approximation used in $H(t)$ above.

\subsection{Equidistant quantum dots}

In the case that the QDs are equidistant from each other, i.e., $N=2$ dots
on a line, $N=3$ dots at the vertices of an equilateral triangle, the
interdot Coulomb interaction $V_{nn^{\prime }}=V$ is independent of $n$ and $%
n^{\prime }$. Thus the spatial symmetry of the Hamiltonian (1) enables us to
introduce the global angular momentum operators\cite{Qui} 
\begin{equation}
J_x=\frac 12\sum_{n=1}^N(e_n^{+}h_n^{+}+h_ne_n),  \eqnum{2a}
\end{equation}
\begin{equation}
J_y=\frac{-i}2\sum_{n=1}^N(e_n^{+}h_n^{+}-h_ne_n),  \eqnum{2b}
\end{equation}
\begin{equation}
J_z=\frac 12\sum_{n=1}^N(e_n^{+}e_n-h_nh_n^{+}),  \eqnum{2c}
\end{equation}
which obey standard angular momentum commutation relationships $[J_\alpha
,J_\beta ]=iJ_\gamma $, where $(\alpha ,\beta ,\gamma )$ represent a cyclic
permutation of $(x,y,z)$. In terms of these new operators the Hamiltonian
for the equidistant QDs can be rewritten as a direct sum over various $J$%
-invariant Hamiltonian, i.e., 
\begin{equation}
H(t)=\oplus _{J=0}^{N/2}H^{(J)}(t),  \eqnum{3}
\end{equation}
where 
\begin{equation}
H^{(J)}(t)=\varepsilon J_z-V(J^2-J_z^2)+\frac 12\Omega (t)e^{-i\omega
t}J_{+}+\frac 12\Omega ^{*}(t)e^{-i\omega t}J_{-},  \eqnum{4}
\end{equation}
where $J_{\pm }=J_x\pm iJ_y$ are the usual raising and lowering operators.
To proceed we introduce the time dependent unitary transformation $U=\exp
(-i\omega tJ_z)$. The transformed Hamiltonian in the rotating frame is 
\begin{equation}
H_{RF}^{(J)}=\Delta J_z-V(J^2-J_z^2)+\Omega _x(t)J_x+\Omega _y(t)J_y, 
\eqnum{5}
\end{equation}
where $\Delta $ is the detuning from resonant excitation, $\Omega _x(t)=%
\mathop{\rm Re}%
(\Omega (t))$ and $\Omega _y(t)=%
\mathop{\rm Im}%
(\Omega (t))$ are the Rabi coupling strength along the $x$ and $y$ axises,
respectively.

\subsection{Quantum dots with a linear configuration}

When the quantum dots are prepared along a ray, the value of $V_{n,n^{\prime
}}$ depends on the $n$ or $n^{\prime }$. Here we assume that the exciton
transfer can only be excited by the hopping between the nearest neighbors..
Thus\ only $V_{n,n+1}=V$ ($n=1,2,...,N-1$) is not zero, while the other $%
V_{nn^{\prime }}$ are zero in the tight-binding approximation. In this case,
we introduce the local $1/2$ speudospin operators 
\begin{equation}
\sigma _n^x=\frac 12(e_n^{+}h_n^{+}+h_ne_n),  \eqnum{6a}
\end{equation}
\begin{equation}
\sigma _n^y=\frac{-i}2(e_n^{+}h_n^{+}-h_ne_n),  \eqnum{6b}
\end{equation}
\begin{equation}
\sigma _n^z=\frac 12(e_n^{+}e_n-h_nh_n^{+}),  \eqnum{6c}
\end{equation}
which obey the commutation relationships among three Pauli matrices $[\sigma
_n^\alpha ,\sigma _{n^{\prime }}^\beta ]=i\delta _{n,n^{\prime }}\sigma
_n^\gamma $. The Hamiltonian can be rewritten in terms of these local 1/2
spin operators as 
\begin{eqnarray}
H(t) &=&\varepsilon \sum_n^N\sigma _n^z-V\sum_{n=1}^{N-1}(\sigma
_n^{-}\sigma _{n+1}^{+}+\sigma _n^{+}\sigma _{n+1}^{-})  \eqnum{7} \\
&&+\frac 12\Omega (t)e^{-i\omega t}\sum_{n=1}^N\sigma _n^{+}+\frac 12\Omega
^{*}(t)e^{i\omega t}\sum_{n=1}^N\sigma _n^{-},  \nonumber
\end{eqnarray}
where $\sigma _n^{\pm }=\sigma _n^x\pm i\sigma _n^y$. In deriving Eq. (7) we
have neglected all constant energy terms which have no contribution to the
dynamics. Again, we transform the Hamiltonian (7) into the rotating frame by
introducing the unitary transformation $U=\exp (-i\omega
_{las}t\sum_n^N\sigma _n^z)$ as follows 
\begin{equation}
H_{RF}=\sum_n^N\Delta \sigma _n^z-V\sum_{n=1}^{N-1}(\sigma _n^{-}\sigma
_{n+1}^{+}+\sigma _n^{+}\sigma _{n+1}^{-})+\Omega (t)\sum_{n=1}^N\sigma
_n^{+}+\Omega ^{*}(t)\sum_{n=1}^N\sigma _n^{-}.  \eqnum{8}
\end{equation}
In the absence of optical field, the Hamiltonian (8) is identical to an
one-dimensional X-Y model in the magnetic system. In the limit $N\rightarrow
\infty $, one can obtain the exact ground state with the help of well-known
Jordan-Wigner transformation.

\section{Bell state generation in double QDs}

To give a systematic analysis on the exciton dynamics, we start with the
exploitation of the maximally entangled Bell state generation. In the
absence of optical excitation, there is no interband transition, so there is
no excitons in the double QDs, i.e., we start with the vacuum state $%
|00\rangle $. In the following we will show how to generate the maximally
entangled Bell state of the form $|\Psi _{Bell}\rangle =(1/\sqrt{2}%
)(|00\rangle +e^{i\phi }|11\rangle )$ with $0$ ($1$) denoting a zero-exciton
(single-exciton) QD. According to Eq. (2), the initial vacuum state $%
|00\rangle $ is identical to $|J=1,J_z=-1\rangle $ (denoted by $|1,-1\rangle 
$ in the following) in the angular momentum representation, thus the
subsequent time evolution in the presence of the laser field will be
restricted to the $J=1$ subspace. This means that the antisymmetric
single-exciton state is light-inactive.. The evolution of any initial state $%
\mid \Psi (0)\rangle $ under the action of $H_{RF}^{(J=1)}$ in Eq. (5) can
be thus expressed as $|\Psi (t)\rangle =c_1(t)|1,1\rangle +c_2(t)|1,0\rangle
+c_3(t)|1,-1\rangle $ in the angular momentum representation. Here the
coefficients $c_k(t)$ are determined by the Schr\"{o}dinger equation 
\begin{equation}
i\left( 
\begin{array}{l}
\stackrel{.}{c}_1 \\ 
\stackrel{.}{c}_2 \\ 
\stackrel{.}{c}_3
\end{array}
\right) =\left( 
\begin{array}{lll}
\Delta -V & |\Omega |e^{-i\varphi }/\sqrt{2} & 0 \\ 
|\Omega |e^{i\varphi }/\sqrt{2} & -2V & |\Omega |e^{-i\varphi }/\sqrt{2} \\ 
0 & |\Omega |e^{i\varphi }/\sqrt{2} & -\Delta -V
\end{array}
\right) \left( 
\begin{array}{l}
c_1 \\ 
c_2 \\ 
c_3
\end{array}
\right) ,  \eqnum{9}
\end{equation}
where $|\Omega |=\sqrt{\Omega _x^2+\Omega _y^2}/2$ and $\varphi =\tan
^{-1}(\Omega _y/\Omega _x)$. Therefore the probability $\rho _{Bell}$ for
finding the maximally entangled Bell state in a double quantum dot is given
by

\begin{equation}
\rho _{Bell}=\frac 12\left| c_3(t)+e^{i\phi }c_1(t)\right| ^2.  \eqnum{10}
\end{equation}

The eigenenergies associated with the Schr\"{o}dinger equation (9) can be
solved analytically for general values of driving frequency $\omega $. For
brevity we do not give the explicit expressions here. Instead we illustrate
in Fig. 1 the spectrum features by plotting the eigenenergies as a function
of driving frequency. It shows in Fig. 1 that an avoided crossing between
energies $E_1$ and $E_2$ occurs at the value of $\omega =\varepsilon $,
which corresponds to the exact resonance condition $\Delta =0$. The
occurrence of avoided crossing in the energy spectrum implies the strong
resonant oscillation between the corresponding eigenstates. The oscillation
frequency can be easily read out from the difference between the energy
levels at $\Delta =0$. In this case the eigenenergies and eigenstates (not
normalized) of Eq. (9) are

\begin{eqnarray}
&\mid &\varphi _1\rangle =|1,1\rangle -\frac b{\sqrt{2}|\Omega |}|1,0\rangle
+|1,-1\rangle ,\text{ \ \ \ \ \ \ \ \ \ \ \ \ \ \ \ \ \ }E_1=a-V, 
\eqnum{11a} \\
&\mid &\varphi _2\rangle =-|1,1\rangle +|1,-1\rangle \text{ \ \ \ \ \ \ \ \
\ \ \ \ \ \ \ \ \ \ \ \ \ \ \ \ \ \ \ \ \ \ \ \ \ \ \ \ \ \ \ }E_2=-2V, 
\eqnum{11b} \\
&\mid &\varphi _3\rangle =|1,1\rangle -\frac a{\sqrt{2}|\Omega |}|1,0\rangle
+|1,-1\rangle \text{ \ \ \ \ \ \ \ \ \ \ \ \ \ \ \ \ \ }E_3=b-V,  \eqnum{11c}
\end{eqnarray}
where $a=(-V-\sqrt{V^2+4|\Omega |^2})/2$, and $b=(-V+\sqrt{V^2+4|\Omega |^2}%
)/2$. From Eq. (11) we can see that for weak driving field $|\Omega |\ll V$,
the states $|\varphi _1\rangle $ and $\mid \varphi _2\rangle $ are nearly
degenerate and dominated by the zero exciton state $|1,-1\rangle $ and
double exciton state $\mid 1,1\rangle $, whereas the state $|\varphi
_3\rangle $ is dominated by the single exciton state $\mid 1,0\rangle $.
Starting from the initial state $|1,-1\rangle $, we expect its resonant
oscillation with $|1,1\rangle $, with the oscillation frequency approximated
by

\begin{equation}
\omega _r=E_2-E_1\simeq |\Omega |^2/V.  \eqnum{12}
\end{equation}
Because the population of the single-exciton state remains very small during
time evolution, we can approximate $c_2(t)$ in Eq. (9) to first order of $%
|\Omega |/V$

\begin{equation}
c_2(t)=\frac{|\Omega |}{\sqrt{2}V}e^{i\varphi }c_1(t)+\frac{|\Omega |}{\sqrt{%
2}V}e^{-i\varphi }c_3(t).  \eqnum{13}
\end{equation}
By introducing $c_2(t)$ from Eq. (13) in the Schr\"{o}dinger equation we
reduce the system to an effective two-level system. The reduced equation has
the form

\begin{equation}
i\left( 
\begin{array}{l}
\stackrel{.}{c}_1 \\ 
\stackrel{.}{c}_3
\end{array}
\right) =\left( 
\begin{array}{ll}
-V+\frac{|\Omega |^2}{2V} & \frac{|\Omega |^2}{2V}e^{-i2\varphi } \\ 
\frac{|\Omega |^2}{2V}e^{i2\varphi } & -V+\frac{|\Omega |^2}{2V}
\end{array}
\right) \left( 
\begin{array}{l}
c_1(t) \\ 
c_3(t)
\end{array}
\right) .  \eqnum{14}
\end{equation}
Thus with the initial zero-exciton state, we have the time evolution of the
system as follows

\begin{eqnarray}
c_1(t) &=&-i\exp [i(V+\frac{|\Omega |^2}{2V})t)\exp (-i2\varphi )\sin
[|\Omega |^2t/(2V)]  \eqnum{15a} \\
c_3(t) &=&\exp [i(V+\frac{|\Omega |^2}{2V})t)\cos [|\Omega |^2t/(2V)] 
\eqnum{15b}
\end{eqnarray}
Substituting Eq. (15) into Eq. (10) we have the probability for finding the
Bell state $(1/\sqrt{2})(|00\rangle +e^{i\phi }|11\rangle )$ at time $t$

\begin{equation}
\rho _{Bell}(t)=\frac 12[1+\sin (\omega _rt)\cos (\phi -2\varphi -\pi /2)], 
\eqnum{16}
\end{equation}
where $\omega _r=|\Omega |^2/V$. From Eq. (16) one can see that the Bell
state with arbitrary phase can be generated by controlling the Rabi coupling
strength. In the case of $\Omega _y=0$ and constant value of $\Omega _x$, we
obtain the same result as in Ref. [13]. Note that the Bell-state generation
time is significantly shortened by applying stronger laser pulses. This is
important because short pulse of length for Bell-state generation is
fundamental to experimental observation of such maximally entangled state,
which is impeded by inevitable decoherence occurred in the realistic double
quantum dot system. We find Eq. (16) is remarkably valid for the slowly
varying amplitude $\Omega (t)$.

For numerical calculations, we consider Gaussian temporal pulse shape for
the excitation laser. The time-dependent Schr\"{o}dinger equation is
numerically integrated using the fourth order Runge-Kutta scheme. The
results of the Bell-state generation dynamics are shown in Fig. 2. The laser
pulse shape is plotted as a dotted line. The square amplitudes of the vacuum
state $|00\rangle $ and biexciton state $|11\rangle $ are denoted by $%
|c_3|^2 $ and $|c_1|^2$, respectively, and plotted as solid lines. The
population of single-exciton state is given by $|c_2|^2$. As one can see
from Fig. 2, the quantity of $|c_2|^2$ is always near zero during time
evolution. This light-inactive property enables us to adiabatically
eliminate its contribution and reduce the system to an effective two-level
model, as we have done in deriving Eq. (16). The probability $\rho _{Bell}$
for finding the maximally entangled Bell state $(1/\sqrt{2})(|00\rangle
+e^{i\pi /2}|11\rangle )$ is also shown in the figure as dashed line. It
achieves its maximum value of almost unity in the middle of optical
excitation and remains unchanged afterwards.

\section{GHZ state generation in equidistant quantum dots}

In this section we show the optical excitation of maximally entangled GHZ
state $(1/\sqrt{2})(|000\rangle +e^{i\phi }|111\rangle )$ in three coupled
QDs with equidistance from each other. The initial state is chosen to be the
vacuum state $|000\rangle $, i.e., the eigenstate $|3/2,-3/2\rangle $ of the
angular momentum operator $J_z$. Thus the subsequent time evolution of the
system is confined to the $J=3/2$ subspace. The evolution of wave function
can be expressed as 
\[
|\Psi (t)\rangle =c_1(t)|3/2,3/2\rangle +c_2(t)|3/2,1/2\rangle
+c_3(t)|3/2,-1/2\rangle +c_4(t)|3/2,-3/2\rangle . 
\]
where the coefficients $c_k(t)$ are determined by the Schr\"{o}dinger
equation 
\begin{equation}
i\frac d{dt}\left( 
\begin{array}{l}
c_1 \\ 
c_2 \\ 
c_3 \\ 
c_4
\end{array}
\right) =\left( 
\begin{array}{llll}
\frac{3(-V+\Delta )}2 & \sqrt{3}|\Omega |e^{-i\varphi }/2 & 0 & 0 \\ 
\sqrt{3}|\Omega |e^{i\varphi }/2 & \frac{-7V+\Delta }2 & |\Omega
|e^{-i\varphi } & 0 \\ 
0 & |\Omega |e^{-i\varphi } & \frac{-7V-\Delta }2 & \sqrt{3}|\Omega
|e^{-i\varphi }/2 \\ 
0 & 0 & \sqrt{3}|\Omega |e^{i\varphi }/2 & \frac{3(-V-\Delta )}2
\end{array}
\right) \left( 
\begin{array}{l}
c_1 \\ 
c_2 \\ 
c_3 \\ 
c_4
\end{array}
\right) .  \eqnum{17}
\end{equation}
The probability for finding the maximally entangled GHZ state is given by 
\begin{equation}
\rho _{GHZ}=\frac 12\left| c_4(t)+e^{i\phi }c_1(t)\right| ^2.  \eqnum{18}
\end{equation}

Figure 3 shows the eigenenergy spectrum of the Hamiltonian in Eq. (17) as a
function of the driving frequency. It shows that there are two avoided
crossings in the energy spectrum when the driving frequency approaches to
satisfy the resonance condition $\Delta =0$, which implies resonant
oscillations between the relevant eigenstates. The oscillation frequency can
be obtained from the difference of the energy levels at $\Delta =0$. In this
case, the eigenenergies are

\begin{equation}
E_{1,2}=-5V/2\mp |\Omega |-\sqrt{(-V\mp |\Omega |)^2+3|\Omega |^2}, 
\eqnum{19a}
\end{equation}
\begin{equation}
E_{3,4}=-5V/2\mp |\Omega |+\sqrt{(-V\mp |\Omega |)^2+3|\Omega |^2}. 
\eqnum{19b}
\end{equation}
The corresponding un-normalized eigenstates are given by 
\begin{equation}
|\varphi _i\rangle =|3/2,-3/2\rangle +\delta _i|3/2,-1/2\rangle \mp \delta
_i|3/2,1/2\rangle +|3/2,3/2\rangle \text{ }(i=1,...,4),  \eqnum{20}
\end{equation}
where $\delta _i=(E_i+3V/2)/(\sqrt{3}|\Omega |)$. From Eqs. (19)-(20) one
can see that for weak driving field $|\Omega |\ll V$, the states $|\varphi
_3\rangle $ and $|\varphi _4\rangle $ are nearly degenerate and dominated by
the vacuum state $|3/2,-3/2\rangle $ and triexciton state $|3/2,3/2\rangle $%
, whereas the states $|\varphi _1\rangle $ and $|\varphi _2\rangle $ are
dominated by the single-exciton state $|3/2,-1/2\rangle $ and biexciton
state $|3/2,1/2\rangle $. Thus under the initial vacuum state condition, the
dynamic evolution of the system is characterized by the resonant oscillation
between $|3/2,-3/2\rangle $ and $|3/2,3/2\rangle $, whereas the contribution
from the states $|3/2,-1/2\rangle $ and $|3/2,1/2\rangle $ can be neglected.
To adiabatically eliminate these two states from the dynamics, we introduce
the unitary transformation 
\begin{equation}
R=\left( 
\begin{array}{llll}
1 & 0 & 0 & 0 \\ 
0 & \frac 1{\sqrt{2}} & \frac 1{\sqrt{2}} & 0 \\ 
0 & \frac 1{\sqrt{2}}e^{i\varphi } & -\frac 1{\sqrt{2}}e^{i\varphi } & 0 \\ 
0 & 0 & 0 & 1
\end{array}
\right) ,  \eqnum{21}
\end{equation}
which transforms the state components $c_2(t)$ and $c_3(t)$ into the
diagonal representation. Defining the state vector $\overrightarrow{c}%
=(c_1,...,c_4)^T$, supposing $\overrightarrow{c}(t)=e^{i3Vt/2}R%
\overrightarrow{d}(t)$, we obtain the equation of motion for the reduced
state vector 
\begin{equation}
i\frac d{dt}\overrightarrow{d}(t)=H_d^{(J=3/2)}\overrightarrow{d}(t), 
\eqnum{22}
\end{equation}
where the transformed resonant Hamiltonian ($\Delta =0$) is 
\begin{eqnarray}
H_d^{(J=3/2)} &=&R^{+}H_{RF}^{(J=3/2)}R+3V/2  \nonumber \\
&=&\left( 
\begin{array}{llll}
0 & \sqrt{\frac 38}|\Omega |e^{-i\varphi } & \sqrt{\frac 38}|\Omega
|e^{-i\varphi } & 0 \\ 
\sqrt{\frac 38}|\Omega |e^{i\varphi } & -2V+|\Omega | & 0 & \sqrt{\frac 38}%
|\Omega |e^{-i2\varphi } \\ 
\sqrt{\frac 38}|\Omega |e^{i\varphi } & 0 & -2V-|\Omega | & -\sqrt{\frac 38}%
|\Omega |e^{-i2\varphi } \\ 
0 & -\sqrt{\frac 38}|\Omega |e^{i2\varphi } & -\sqrt{\frac 38}|\Omega
|e^{i2\varphi } & 0
\end{array}
\right) .  \eqnum{23}
\end{eqnarray}
The two components $d_2$ and $d_3$ can now be adiabatically eliminated in
the same matter in deriving Eq. (15). Thus one obtains the effective
two-state approximation as follows 
\begin{equation}
i\frac d{dt}d_1(t)=\chi _1d_1(t)+e^{-i3\varphi }\chi _2d_4(t),  \eqnum{24a}
\end{equation}
\begin{equation}
i\frac d{dt}d_4(t)=e^{i3\varphi }\chi _2d_1(t)+\chi _1d_4(t),  \eqnum{24b}
\end{equation}
where 
\begin{equation}
\chi _1=\frac{3|\Omega |^2}{16V-8|\Omega |}+\frac{3|\Omega |^2}{16V+8|\Omega
|},  \eqnum{25a}
\end{equation}
\begin{equation}
\chi _2=\frac{3|\Omega |^2}{16V-8|\Omega |}-\frac{3|\Omega |^2}{16V+8|\Omega
|}.  \eqnum{25b}
\end{equation}
With the initial condition $d_1(0)=0$ and $d_4(0)=1$, one obtains the
solution of Eq. (24) 
\begin{equation}
d_1(t)=-ie^{-i3\varphi }e^{i\chi _1t}\sin (\chi _2t),  \eqnum{26a}
\end{equation}
\begin{equation}
d_4(t)=e^{i\chi _1t}\cos (\chi _2t).  \eqnum{26b}
\end{equation}
Substituting Eqs. (26) into the expression for $\rho _{GHZ}(t)$, one obtains
the probability for finding the maximally entangled GHZ state $|\Psi
_{GHZ}\rangle $ 
\begin{equation}
\rho _{GHZ}(t)=\frac 12\{1+\sin (\omega _rt)\cos (\phi -3\varphi -\pi /2)\},
\eqnum{27}
\end{equation}
where the oscillating frequency $\omega _r=2\chi _2\simeq 3|\Omega
|^3/(8V^2) $. Eq. (27) shows that the maximally entangled GHZ state with
arbitrary phase can be generated by a selective pulse of laser field. In
particular, in the case of $\Omega _y=0$, a $\pi /2$ pulse produces the GHZ
state $(|000\rangle +e^{i\pi /2}|111\rangle )/\sqrt{2}$ at time $\tau
_G=4\pi V^2/(3|\Omega |^3)$. Note that the result of Eq. (27) in the case of 
$\varphi =0$ was first obtained by Quiroga and Johnson in the density matrix
formalism\cite{Qui}. Our approach, which is based on a combination of
eigenenergy spectrum analysis and adiabatic elimination of dark states, may
be combined with the density matrix method to highlight the physical
prospects in preparing entangled qubits.

To compare the analytical and numerical solutions for the unitary evolution
described above. We show in Fig. 4 the time evolution of $\rho _{GHZ}(t)$
with $\Omega _y=0$ and a constant value of $\Omega _x=0.2V$. The solid line
in Fig. 4 is the exact solution by numerically integrating Schr\"{o}dinger
equation (17), whereas the dotted line is the result of Eq. (27). Clearly
our two-state approximation describes the system's evolution very well when
compared with the exact numerical solution, implying that the system's
quantum state at time $\tau _G$ corresponds to a maximally entangled GHZ
state $|\Psi _{GHZ}\rangle =(|000\rangle +e^{i\pi /2}|111\rangle )/\sqrt{2}$%
. For a more realistic consideration, we employ Gaussian temporal pulse
shape and present in Fig. 5 the generation process of the GHZ state $|\Psi
_{GHZ}\rangle $. Again, one can see that the dynamics of the system is
dominated by the entanglement of the vacuum state and triexciton state,
while the population of single- and bi-exciton states are strongly
suppressed. As a consequence, the probability $\rho _{GHZ}$ achieves and
remains unity after laser pulse.

\section{GHZ state generation in a ray of three coupled QDs}

In this section we show the optical excitation of maximally entangled GHZ
states in a ray of three coupled QDs. The dynamics of the system is now
described by the Hamiltonian (8). In an exciton number basis consisting of $%
|000\rangle $, $|100\rangle $, $|010\rangle $, $|001\rangle $, $|110\rangle $%
, $|011\rangle $, $|101\rangle $, and $|111\rangle $, the Schr\"{o}dinger
equation is 
\begin{equation}
i\frac d{dt}\left( 
\begin{array}{l}
c_1 \\ 
c_2 \\ 
c_3 \\ 
c_4 \\ 
c_5 \\ 
c_6 \\ 
c_7 \\ 
c_8
\end{array}
\right) =\left( 
\begin{array}{llllllll}
-3\Delta & \Omega ^{*} & \Omega ^{*} & \Omega ^{*} & 0 & 0 & 0 & 0 \\ 
\Omega & -\Delta & -V & 0 & \Omega ^{*} & 0 & \Omega ^{*} & 0 \\ 
\Omega & -V & -\Delta & -V & \Omega ^{*} & \Omega ^{*} & 0 & 0 \\ 
\Omega & 0 & -V & -\Delta & 0 & \Omega ^{*} & \Omega ^{*} & 0 \\ 
0 & \Omega & \Omega & 0 & \Delta & 0 & -V & \Omega ^{*} \\ 
0 & 0 & \Omega & \Omega & 0 & \Delta & -V & \Omega ^{*} \\ 
0 & \Omega & 0 & \Omega & -V & -V & \Delta & \Omega ^{*} \\ 
0 & 0 & 0 & 0 & \Omega & \Omega & \Omega & 3\Delta
\end{array}
\right) \left( 
\begin{array}{l}
c_1 \\ 
c_2 \\ 
c_3 \\ 
c_4 \\ 
c_5 \\ 
c_6 \\ 
c_7 \\ 
c_8
\end{array}
\right) .  \eqnum{28}
\end{equation}
The probability for finding the GHZ state $|\Psi _{GHZ}\rangle =(|000\rangle
+e^{i\phi }|111\rangle )/\sqrt{2}$ is given by $\rho
_{GHZ}(t)=|c_8(t)+e^{i\phi }c_1|^2/2$.

Without knowing an analytical approximation of Eq. (28), we turn to
numerically show the optical excitation of the GHZ state. In the absence of
laser field, one can see from Eq. (28) that the subspaces of vacuum,
exciton, biexciton, and triexciton states are not coupled. In this case, the
typical energy spectrum is shown in Fig. 6(a) as a function of detuning $%
\Delta $. It shows in Fig. 6(a) that when $\Delta $ approaches to zero, the
spectrum is characterized by three degenerate energies. The degenerate
states with energy $E=0$ consist of vacuum state $|000\rangle $, triexciton
state $|111\rangle $, and a pair of single- and bi-exciton states. The other
two set of degenerate states consist of a pair of single- and double-exciton
states, respectively. The energy spectrum features are greatly changed in
the presence of the optical field, which can be seen from Fig. 6(b). It
reveals that the degeneracies are completely broken and three avoided
crossings develop near $\Delta =0$. Among these crossings, the energy
splitting between the eigenstates dominated by the states $|000\rangle $ and 
$|111\rangle $ is smallest, since these two states are coupled in an
indirect way. Therefore, starting from the state $|000\rangle $, we expect
the subsequent time evolution of the system is featured by the resonant
oscillations between the vacuum and triexciton states. This is numerically
verified in Fig. 7, where Fig. 7(a) plots the probabilities for finding the
system in the zero- and triple-exciton states and Fig. 7(b) the probability $%
\rho _{GHZ}(t)$. Clearly it shows that a selective pulse of laser field can
be used to produce the maximally entangled GHZ states in a ray of three QDs.
Note that compared with the results in Fig. 4, it shows in Fig. 7 that {\it %
the GHZ state generation time for a linear configuration is shorter than for
a equidistant configuration}. Thus the linear configuration of three QDs is
preferred to implement the maximally entangled GHZ states for its shorter
generation time.

\section{Conclusion}

In summary, we have studied the optically-controlled exciton dynamics in
multiple QD systems. We have shown that the robust occurrence of avoided
crossing in the eigenenergy spectrum enables the dynamics to be confined to
a reduced two-state Hilbert space, in which the generation of maximally
entangled Bell states and GHZ states with arbitrary phase can be controlled
by selective pulses of classical coherent optical light. The entangled state
generation time decreases significantly with an increase of the laser pulses
strength. We have also found that the GHZ states can be implemented in a
three QD system with linear configuration, with the generation time much
shorter than in an equidistant configuration. The results are expected to be
useful in exploiting the realizations of entanglement in quantum dot systems.

\begin{center}
{\large ACKNOWLEDGMENT}
\end{center}

This work was partially supported by a grant from the Research Committee of
The Hong Kong Polytechnic University (Grant No. G-T308).

{\Large Figure Captions}

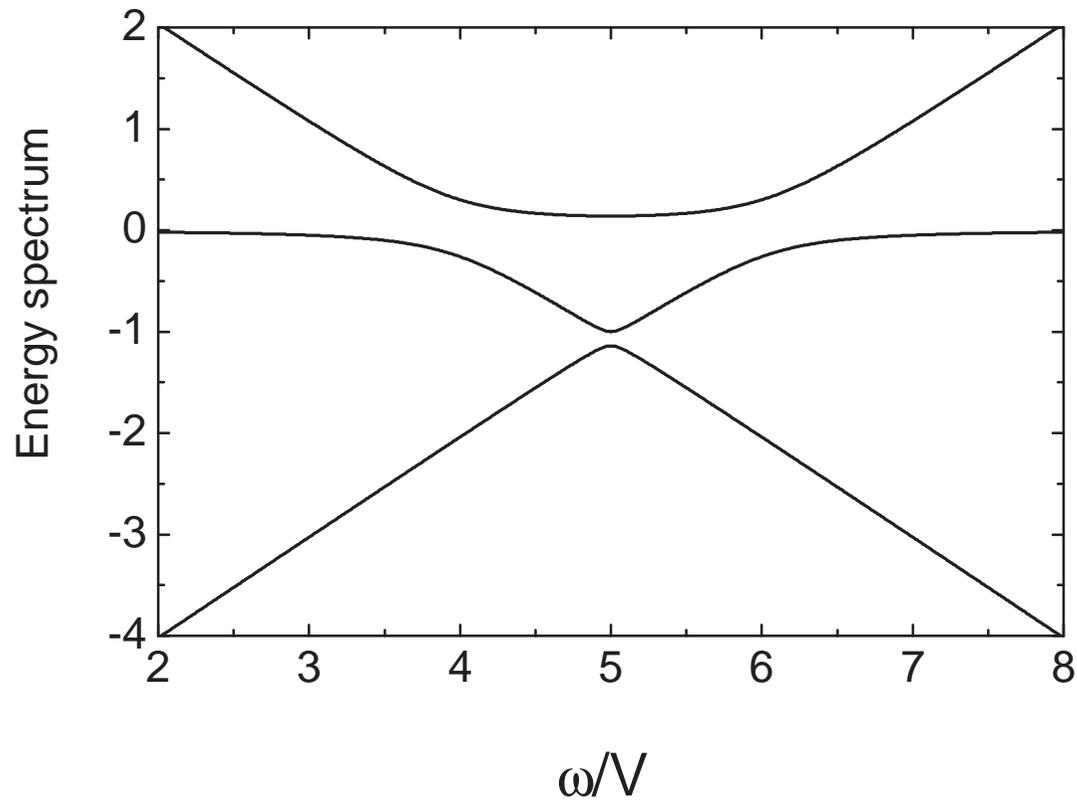
Fig. 1. The energy spectrum of a double quantum dot system as a function of
the frequency of laser pulse. Parameters are $\varepsilon =5V$ and $|\Omega
|=0.2V$.

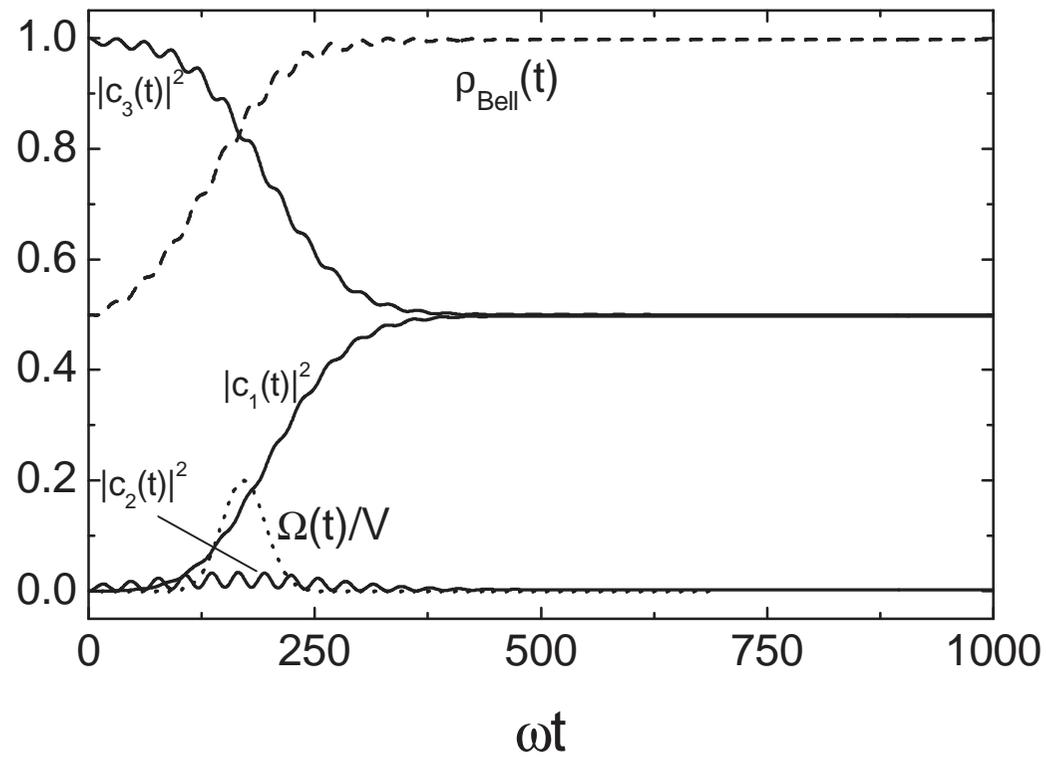
Fig. 2. The Bell-state generation process as a function of time. The pulse
shape $\Omega (t)$ is plotted as a dotted line. The probability $\rho
_{Bell}(t)$ of maximally entangled Bell state is shown as a dashed line. The
population of three exciton number states are also plotted (solid lines).

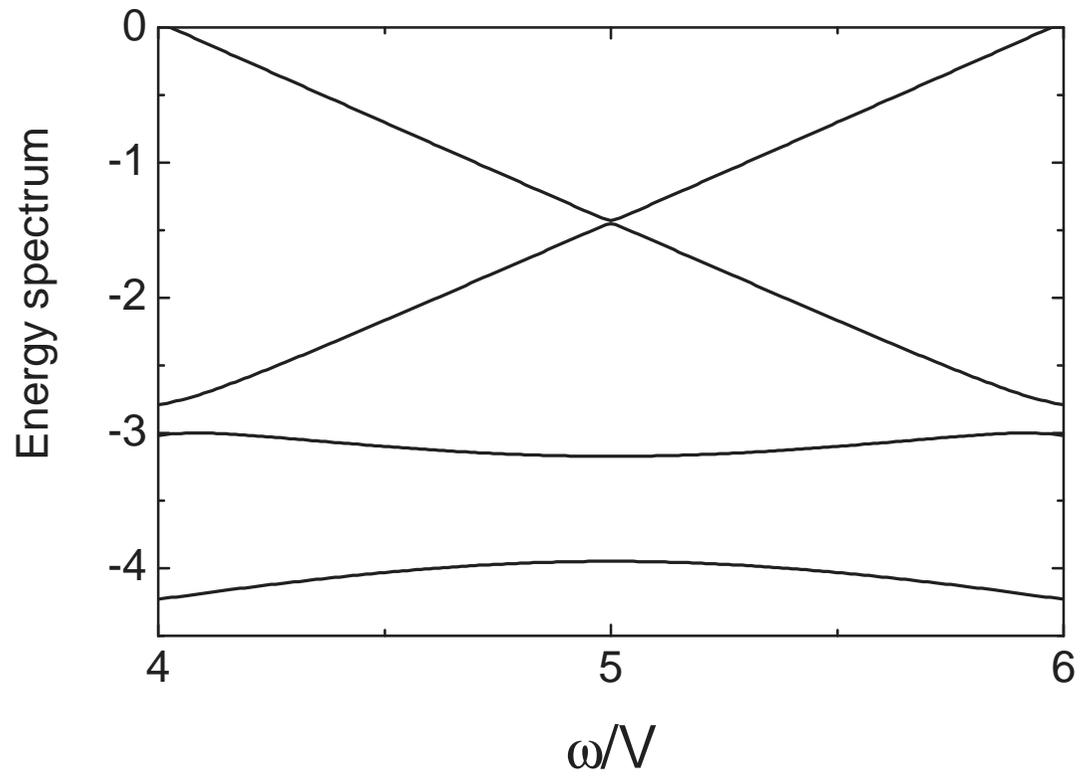
Fig. 3. The energy spectrum of a three quantum dot system as a function of
the frequency of laser pulse. Parameters are $\varepsilon =5V$ and $|\Omega
|=0.2V$.

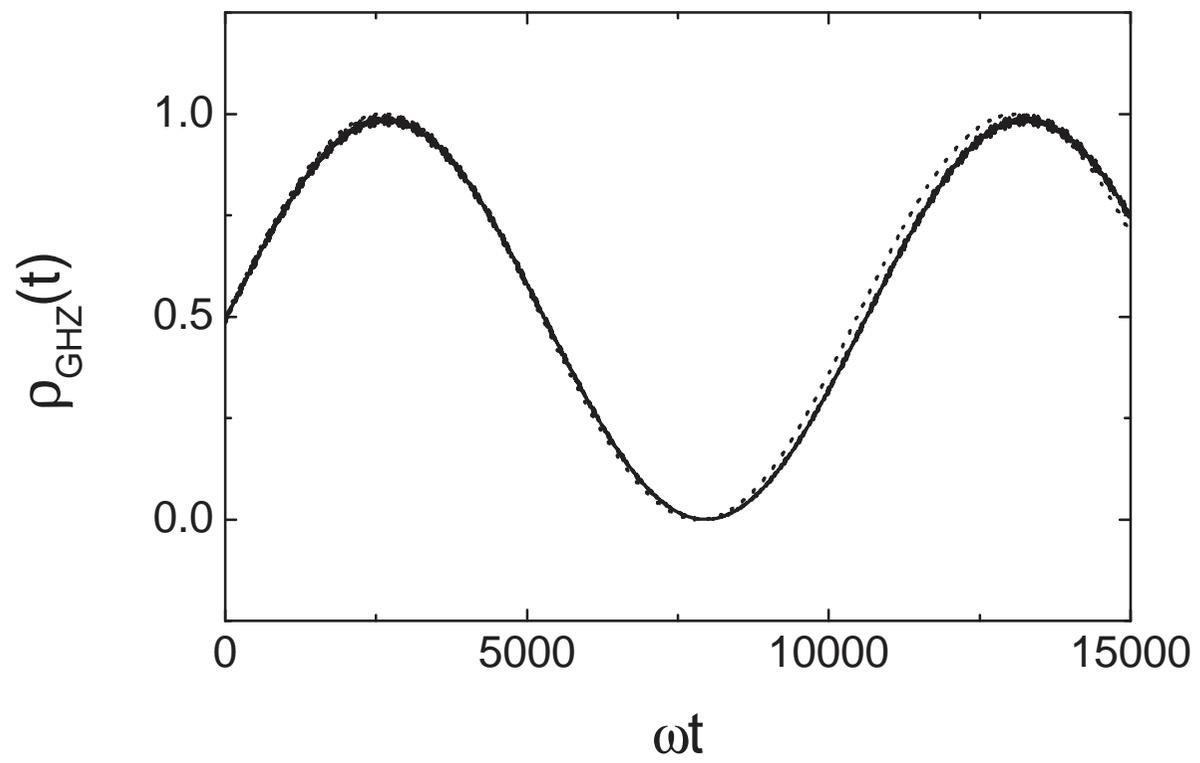
Fig. 4. Exact numerical (solid line) and approximate results of the time
evolution of the probability $\rho _{GHZ}(t)$. Parameters are $\Delta =0$, $%
|\Omega |=0.2V$, and $\varphi =0$.

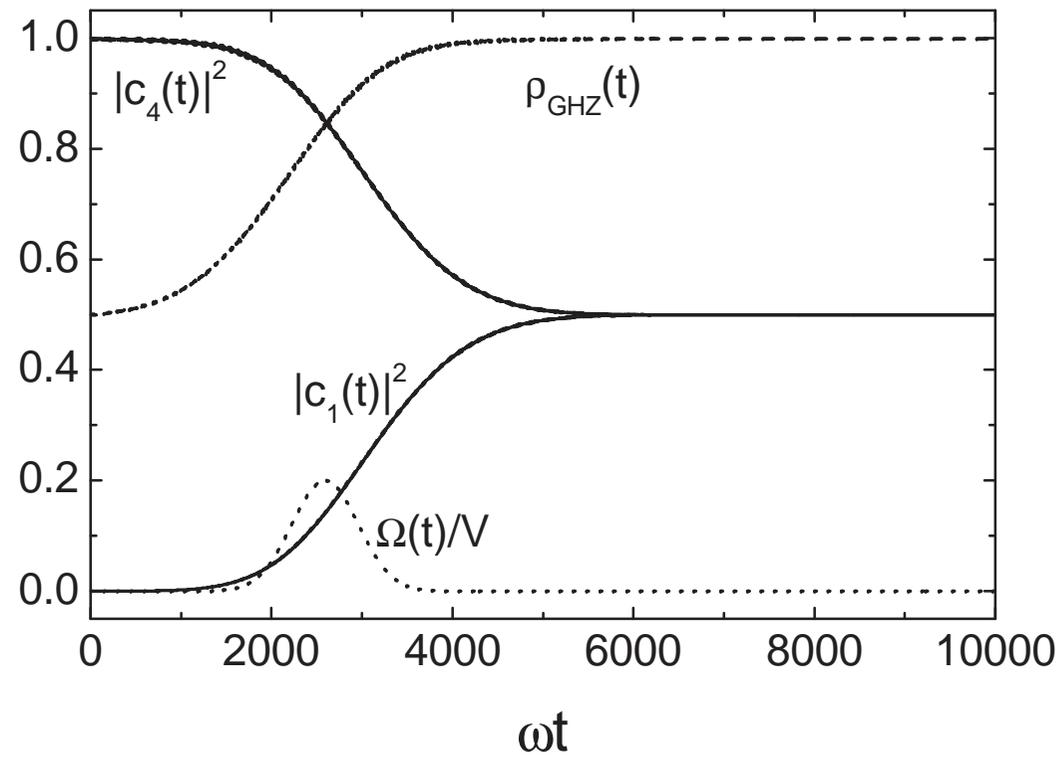
Fig. 5. The GHZ state generation process. The pulse shape $\Omega (t)$ is
plotted as a dotted line. The population of single-exciton and biexciton
states are again strongly suppressed and the probability $\rho _{GHZ}(t)$
(dashed line) is unity after laser pulse.

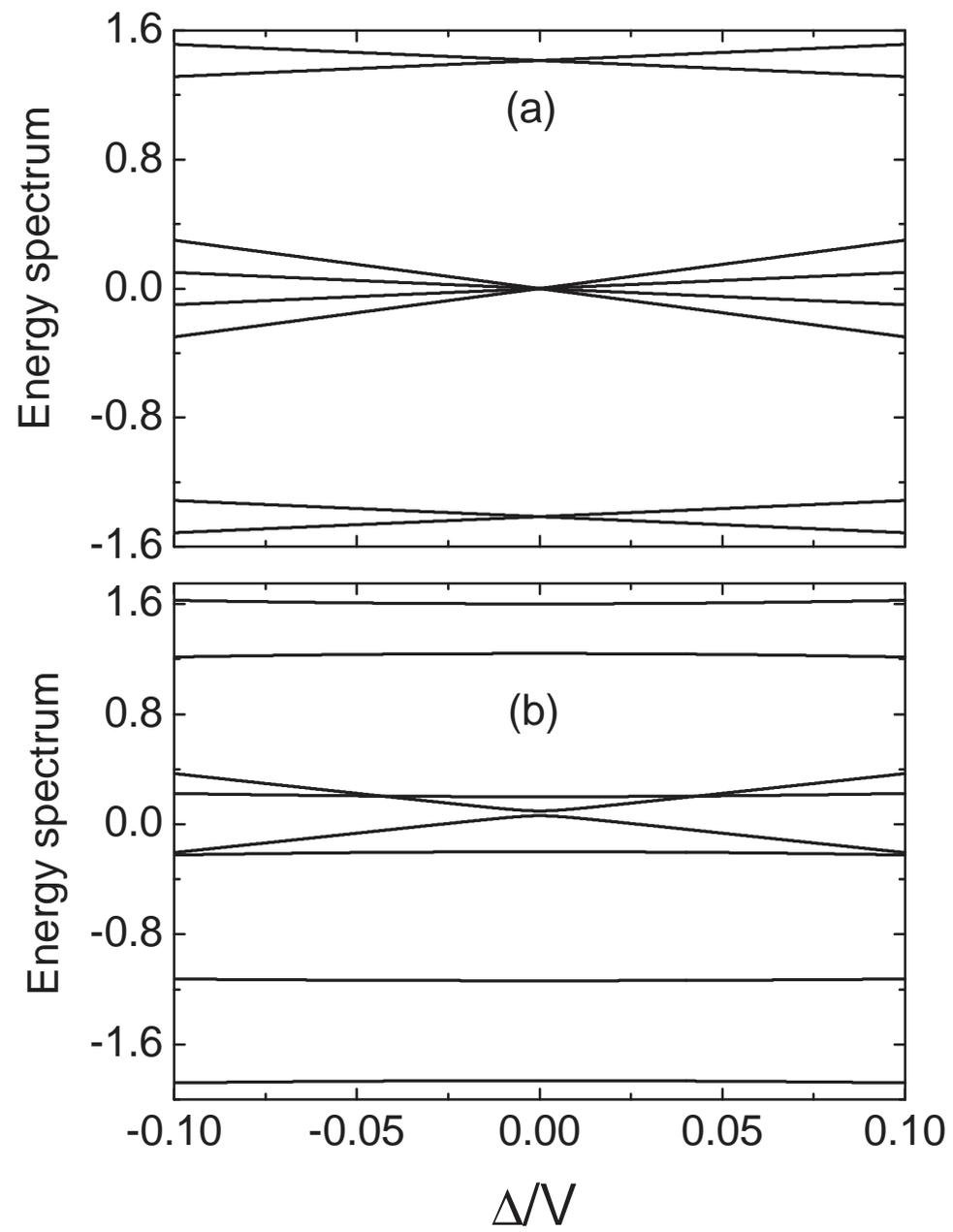
Fig. 6. The energy spectrum of a ray of three quantum dots as a function of
detuning $\Delta $ (a) in the absence of laser field, (b) in the presence of
laser field for the value of $|\Omega |=0.2V$. Other parameter are $%
\varepsilon =5V$.

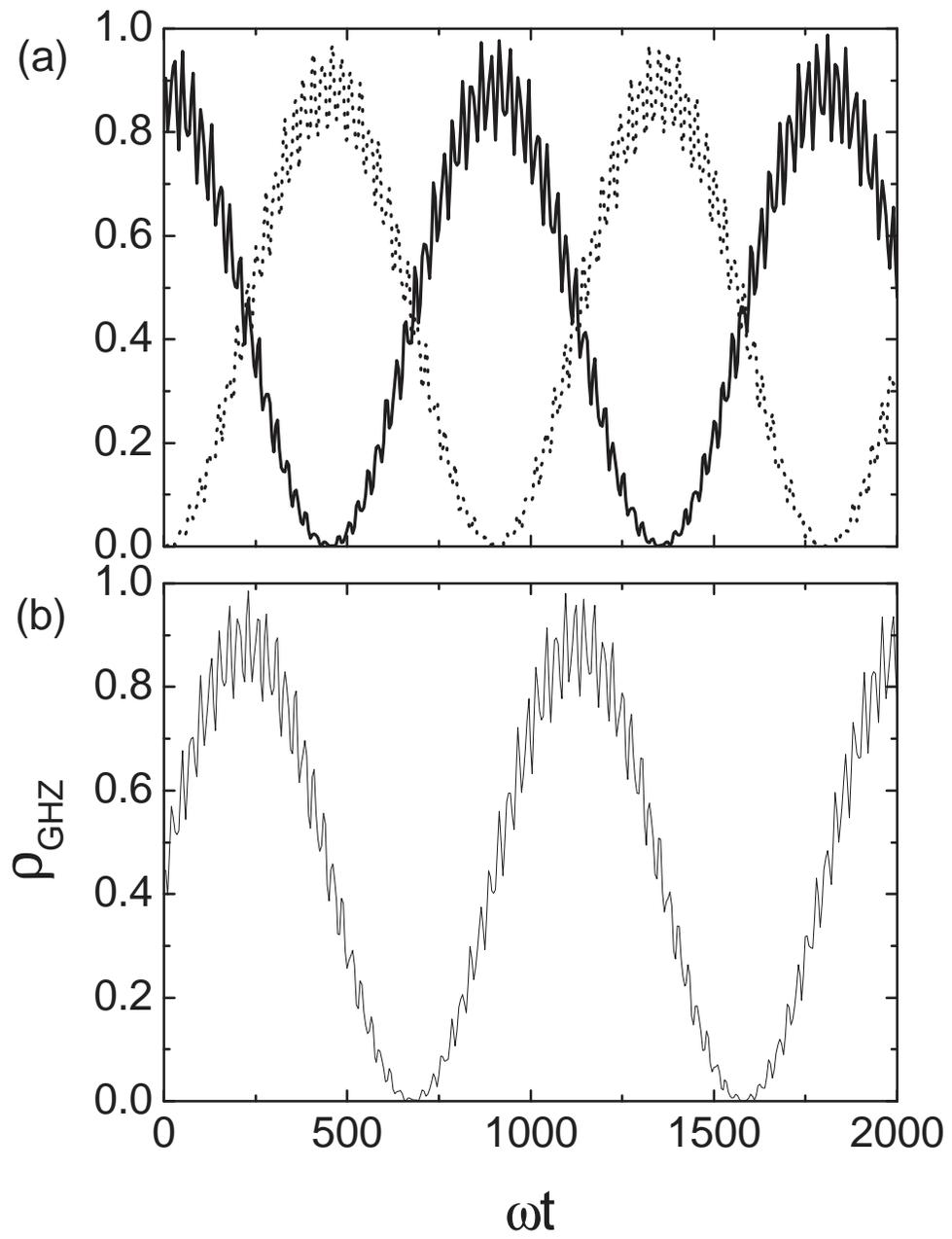
Fig. 7. (a) Time evolution of the population of the vacuum state (solid
line) and triexciton state (dotted line). (b) Time evolution of the
probability $\rho _{GHZ}(t)$ for finding the maximally entangled GHZ state ($%
\phi =\pi /2$) in a ray of three coupled quantum dots. Parameters are $%
\Delta =0$, $\varepsilon =5V$, $|\Omega |=0.2V$, and $\varphi =0$.

\end{document}